\documentstyle[12pt,epsf]{article}

\large
\newcommand{\be}{\begin{eqnarray}}
\newcommand{\ee}{\end{eqnarray}}

\begin{document}
\setlength{\baselineskip}{23pt}
\setlength{\baselineskip}{27pt}
\pagestyle{empty}
\renewcommand{\thefootnote}{\fnsymbol{footnote}}
\centerline{\bf\LARGE Can hadrons survive the}
\centerline{\bf\LARGE chiral phase transition? }
\vskip 1cm
\centerline{\bf T.~Sch\"afer and E.V.~Shuryak\footnote{
Supported in part by the US Department of Energy under
Grant No. DE-FG02-88ER40388.}}
\vskip 1cm
\centerline{\it Department of Physics}
\centerline{\it State University of New York at Stony Brook}
\centerline{\it Stony Brook, New York 11794, USA}
\vskip 1cm

\setlength{\baselineskip}{16pt}
\centerline{\bf Abstract}
   We study mesonic and baryonic correlation functions in the temporal
directions in the vicinity of the critical temperature $T\simeq T_c$ using
the instanton liquid model, in which chiral symmetry restoration
is driven by the formation of instanton-antiinstanton molecules. Although
we find the signals for all hadronic poles to be drastically reduced,
some hadronic states seem to survive the phase transition, as loosely
bound $U(N_f)\times U(N_f)$ chiral multiplets in the plasma phase.
These states are the $0^+ - 0^-$ mesons $(\pi,\sigma,\eta',\delta)$
and the $N(\frac{1}{2}^+) - N(\frac{1}{2}^-)$ baryons.

\vfill
\begin{flushleft}
SUNY-NTG-95-17\\
May 1995
\end{flushleft}
\eject
\newpage
\setlength{\baselineskip}{23pt}
\pagestyle{plain}
\renewcommand{\thefootnote}{\arabic{footnote}}
\setcounter{footnote}{0}
\setcounter{page}{1}

\section{Introduction}

    It was conjectured long ago \cite{CP_75} and later confirmed by a large
number of lattice simulations (see the review \cite{Kar_94}) that at high
temperatures QCD undergoes a transition to a new phase, the quark-gluon
plasma, in which (i) chiral symmetry is restored and (ii) color charges
are {\em screened} \cite{Shu_78} rather than {\em confined}. It is still
unknown whether in the real world (for physical values of quark masses) this
transition is a true singularity or just a rapid crossover. However, it is
well established that qualitative changes happen in a relatively narrow
region $\Delta T \simeq 10\,{\rm MeV} \ll T_c \simeq 150$ MeV. In this
region, for example, the energy density grows by about an order of
magnitude \cite{BKT*95}.

   While physical phenomena below and above this region can (at least
semi-quantitatively) be understood in terms of weekly interacting gases of
hadrons or quark and gluons, respectively, it remains unclear how one should
treat the phenomena close to $T_c$. {\em If} there is no strict phase
transition and {\em if} the time scale is long enough to support a
state in thermodynamic equilibrium, one may expect to see matter
in which the excitation spectrum gradually changes from hadronic states to
quarks and gluons. In terms of spectral densities this means that hadronic
poles gradually dissolve and quark (or gluon) cuts move toward smaller
energies. This is the phenomenon we are going to discuss in this paper.

     For completeness, let us mention two other possibilities.
{\em If} there is a first order transition and the relaxation time is
sufficiently long, a real mixed phase appears, in which the two different
phases coexist. The space and time scale of the domains (or ``bubbles")
is then determined by the production dynamics \cite{CK_92}, mostly
governed by the surface tension $\sigma$.  Available lattice simulations
seem to indicate that it is surprisingly small \cite{KKR_90}. Therfore,
tunneling is relatively easy and one would expect that heavy ion
collisions produce rather {\em well mixed} matter.

   Another special case is the second order transition, in which the
system is governed by the dynamics of long-wavelength fluctuations
of the order parameter. The chiral phase transition in QCD with two
massless quark flavors is such an example, although the real world
might be closer to another second order transition, corresponding to
the boundary of the three flavor first order transition. Whether
or not the critical behavior of the two flavor phase transition
is identical to the O(4) Heisenberg magnet \cite{PW_83} or given by
mean field analysis \cite{KK_95} is currently under debate.

   At zero temperature there is strong support for the idea that
instantons dominate the physics of chiral symmetry breaking and
and explain many properties of light hadrons while confinement plays
a relatively small role \cite{SV_93,CGH*94}. It is quite
plausible that this is also true near the phase transition. At the
moment, there are direct lattice measurements (in quenched QCD) indicating
that the instanton density near the phase transition is very similar
to its value at $T=0$ \cite{CS_95}, in agreement with the arguments
given in \cite{SV_94}. In this letter, we will assume that there is a
sizeable number of instantons present near $T_c$ and study the
implications of this scenario for the survival of hadronic modes
in the ``mixed phase" region.

    The mechanism of the chiral phase transition in the instanton
liquid is related to a rearrangement of the instantons \cite{IS_94,SSV_95},
going from a disordered, random liquid to a phase of correlated
instanton-antiinstanton molecules\footnote{
Similar attempts at understanding the deconfinement transition usually
focus on the rearrangement of monopole loops at $T_c$ \cite{Suz_93}, but
the underlying physics is not very well understood at the moment.}.
In \cite{SSV_95} we have developed a simple model in order
to describe phenomena in the vicinity of the phase transition.
In this model we neglect the finite width $\Delta T$ of the transition
and consider the system at a fixed temperature $T=T_c=150$ MeV. The
relevant parameter in this case is not $T$ but $f$, the fraction of
instantons that are paired in ``molecules". As $f$ is varied from 0
to 1 one passes through the mixed phase region, going from the chirally
broken phase at $f=0$ to the restored phase at $f=1$. All observables
are calculated as a function of the parameter $f$. For example, at
$f\rightarrow 1$ the quark condensate $\langle\bar qq\rangle \rightarrow 0$,
the energy density grows, etc. In \cite{SSV_95} we showed that instantons
generate non-perturbative interactions even when chiral symmetry is restored.
In particular, we were able to reproduce lattice results for spacelike
screening masses near the transition.

    Before we study the behavior of hadrons, we would like to remind the
reader about a few properties of correlation functions at finite temperature
\cite{Shu_93}. Since Lorentz invariance is broken, there are two different
kinds of correlators, which depend either on the spatial distance $x$ or
the temporal separation $\tau$. In the case of spacelike separation one
can go to large $x$ and filter out the lowest exponents known as
``screening masses". Lattice measurements of the screening masses
\cite{TK_87} have revealed an interesting spectroscopy of ``quasi-bound
states". Additional evidence for significant deviations from free quark
propagation was given in \cite{Ber*92}, where the corresponding
``wave functions" were measured. These wave functions are localized
even at very large temperatures, a result that created speculations
about the survival of hadronic bound states at large $T$ \cite{deT_88,Zah_91}.
However, as discussed in detail in \cite{KSB*92,HZ_92}, although these
states indeed exist at {\em any} $T$, they have nothing to do with real
bound states. As $T\rightarrow\infty$ QCD becomes dimensionally
reduced and these ``states" form the spectrum of $d=3$ QCD.

   In order to look for {\em real} bound states, on has to study
{\em temporal} correlation functions. However, at finite temperature
temporal correlators can only be measured for distances $\tau<1/2T$
(about 0.6 fm at $T=T_c$), so one can not trivially filter out the
ground state.  For that reason, temporal correlators
were generally believed to be dominated by perturbative phenomena
and fairly useless for the study of hadronic states: but
with some effort and skill some useful information
about the lowest excitations can still be extracted. On the lattice that
was attempted by Boyd et al.~\cite{BGK*94}, who in particular observed
a dramatic difference between the pion and rho channels.

   In this paper we report our results for a number of temporal correlation
functions in the ``cocktail model" mentioned above. The correlation
functions are calculated from the quark propagator in the instanton liquid.
The details of this  calculations can be found in \cite{SSV_95}. Here
we only note a few important properties of this propagator. First, it
of course satisfies the correct antiperiodic boundary conditions in the
finite temperature euclidean box and reduces to the free propagator at
short distances. Furthermore, we numerically invert the dirac operator
in the basis spanned by the instanton zero modes. This means that the
propagator contains the instanton induced interaction to all orders.
When chiral symmetry is broken, this is the 't Hooft effective interaction,
when the symmetry is restored, it is the effective interaction induced
by molecules discussed in \cite{SSV_95}. In the following, we will study
correlation functions
\be
\label{cor}
\Pi(\tau)&=& <\bar q\Gamma q(\tau)\bar q\Gamma q(0)>,
\ee
where $\bar q\Gamma q$ is a mesonic current with the quantum numbers
of the gamma and isospin matrix $\Gamma$. Correlators with baryonic
quantum numbers are defined analogously. We will normalize the correlation
functions to the corresponding free correlators at the same temperature,
\be
\label{free}
\Pi_0(\tau) &=& {\rm Tr} [\Gamma S_0(T,\tau)\Gamma S_0(T,-\tau)] \, , \\
S_0(T,\tau) &=& {\gamma_0\over 2\pi^2} \sum_{n=-\infty}^{\infty}
 {(-)^n \over (\tau+n/T)^3}  \, .
\ee
If the ratio $R=\Pi(\tau)/\Pi_0(\tau)$ is larger than one,
we will refer to the correlator as attractive, while deviations down
imply repulsive interactions.

    Fig.1a shows the pseudoscalar correlator with the quantum numbers
of the pion. At $f=0.25$ there is a very strong signal which melts down
as $f$ approaches one. Remarkably though, it is {\em not}  proportional
to $(1-f)$ as one might naively expect: in particular, there is a significant
signal at $(1-f)=0.05$. Furthermore, even as chiral symmetry is restored
($f=1$, see fig.2c) the correlator is still larger than the perturbative one,
and the
excess can still be interpreted in terms of a hadronic bound state
supplementing the background from free quark propagation.

    This behaviour can be contrasted to the one shown in fig.1b, for the
scalar-isovector ($\delta$ meson) correlator. At small $f$ one finds
a strongly decaying correlation function, indicative of a strongly repulsive
effective interaction, or the absence of light states in this channel. As
$f\rightarrow 1$ this repulsion disappears, and the the correlator grows
with $f$, until at $f=1$ the correlator is larger than the perturbative
one. Furthermore, one finds that in this case the delta correlation function
is equal to the pion correlator. Since the chiral partner of the pion
is the sigma meson, not the delta, this is a sign of $U_A(1)$ restoration
\cite{Shu_94}. From chiral symmetry one then concludes that all scalar
mesons $(\sigma,\pi,\eta',\delta)$ are degenerate. We have checked this
point by calculating the relevant disconnected diagramms for the $\eta'$
and $\sigma$ correlation functions. Note that $U_A(1)$ is restored not
because the anomaly disappears, but because the formation of molecules
implies that topological charge is screened over very short distances.

     In order to study the properties of the correlation functions in a
more quantitative fashion, we have fitted the mesonic correlators using the
following parametrization
\be
\label{fit}
\Pi(\tau)= \lambda^2 D(T,M,\tau) + {\rm Tr} [\Gamma S^V_m(T,\tau)
\Gamma S^V_m(T,-\tau)]\, ,
\ee
as well as an analogous formula for baryonic correlators.
The first term corresponds to the propagation of a meson resonance
with mass $M$ and coupling constant $\lambda$, while the second term
describes the quark-antiquark continuum. Here, $S^V_m(T,\tau)$ is the
vector (chiral even) part of the temporal propagator of a free quark
with mass $m$ at temperature $T$. As $m$ goes to zero, the quark-antiquark
threshold moves down to zero energy and the correlation function becomes
perturbative.

   Applying this parametrization to the pion correlator we find that
both the resonance mass and the chiral quark mass at $f=1$ are small
(consistent with zero). This allows us to put a lower limit on the
pseudosclar coupling for which we find $\lambda_\pi=1.8\pm 0.3\,
{\rm fm}^{-2}$. The product of $\lambda_\pi$ and the pion decay constant
$f_\pi$ can be determined from the off-diagonal pseudoscalar-axialvector
($\gamma_5$-$\gamma_0\gamma_5$) correlation function. This provides a very
clean measurement, since free quarks do not contribute to this correlator.
{}From the data shown in fig.1c and the coupling constants listed in table 1
one clearly observes how $f_\pi$ goes to the zero as chiral symmetry is
restored while $\lambda_\pi$ remains finite. To study this point in more
detail we have calculated the $f=1$ correlators for various values of the
''valence quark`` mass (the mass that enters the quark propagator). We
find that $\lambda_\pi f_\pi$ scales as the quark mass while $\lambda_\pi$
is finite in the chiral limit. The $f=1$ curves shown in fig.2c are
calculated for $m_u=m_d=4$ MeV, all other curves are for $m_u=m_d=20$ MeV.

   Proceeding from the scalar channels to other cases (vector mesons,
baryons etc.) one knows in advance that for most of them the splitting
between the ground state and excited states is much smaller, and
(at the  distances considered) one finds relatively small deviations
from free quark propagation. At zero temperature one can circumvent
this problem by considering off-diagonal correlators, in which free
quark propagation does not contribute. In addition to the
pseudosclar-axialvector correlator mentioned above, we have also
calculated the vector-tensor correlators $\gamma_\nu$-$\sigma_{\mu\nu}$
(fig.2d).  Again, the correlator can be described nicely in terms of
the propagation of a single rho meson bound state, without a free quark
continuum. For a definition of the corresponding coupling constant
$c_\rho$, see \cite{SV_93}. However, as chiral symmetry is restored ($f=1,m
\rightarrow 0$), this signal disappears. One can check that the
vanishing of the vector-tensor correlator is indeed a consequence
of chiral symmetry restoration, and the small signal that we observe
is consistent with perturbative mass corrections. The diagonal
$\vec\gamma$-$\vec\gamma$ rho meson correlator in the chirally restored
phase is shown in fig.2c. There is no evidence for a resonance. Indeed,
the correlator is slightly suppressed with respect to the perturbative
one. This suppression is consistent with a chiral quark mass $m\simeq 85$
MeV. The diagonal $\gamma_4$-$\gamma_4$ correlator shows a very small
enhancement, numerically smaller but in qualitative agreement with the
molecular interaction discussed in \cite{SSV_95}.

    Let us now proceed to baryon correlators. As explained in detail
in \cite{SV_93}, there are 6 nucleon and 4 delta correlation functions
that can be constructed out of the two Ioffe currents for the nucleon
and the unique delta current. Half of these correlators, $\Pi_{1,3,5}^N$
and $\Pi_{1,3}^\Delta$ in the notation of \cite{SV_93} are chiral odd
and have to vanish as chiral symmetry is restored. $\Pi_{2,4}^N$ as well
as $\Pi_{2,4}^\Delta$ are chiral even and may show resonance signals even
as chiral symmetry is restored. $\Pi_6^N$, the chiral even, off-diagonal
correlator of the two Ioffe currents is special. Chiral symmetry does not
require it to vanish, but since the $U_A(1)$ transformation properties of
the two currents do not match, it is an order parameter for $U_A(1)$
symmetry breaking.

    The chiral odd correlators $\Pi_1^N$ and $\Pi_1^\Delta$ are shown
in figs.2a,c. These correlators receive no contribution from free quark
propagation and one can observe clear resonance signals that
die away as $f\rightarrow 1$. Note that there are several possible
explanations for this behavior. If one includes the lowest postive
and negative parity resonances in the spectral function, then the
correlator $\Pi_1^N$ is proportional to $\beta_{N1}=\lambda^2_{1N(\frac12^+)}
m_{N(\frac12^+)}-\lambda^2_{1N(\frac12^-)}m_{N(\frac12^-)}$, where
$\lambda_{1N(\frac12^+)}$ is the coupling of the first Ioffe current
to the positive parity nucleon, etc. In this expression, we have suppressed
the resonance propagators, and there is an analogous result for the delta.
This means that the correlator can vanish either because the masses go to
zero or the resonances decouple, or because the parity partners become
degenerate and their contributions to the correlator cancel.

    The chiral even nucleon and delta correlators $\Pi_2^{N,\Delta}$
in the chirally restored phase are shown in fig.2c. One observes a
small attractive signal in the nucleon case, while no such signal is
present for the delta. In this case, the resonance coupling is
proportional to $\beta_{N2} = \lambda^2_{1N(\frac12^+)} +
\lambda^2_{1N(\frac12^-)}$, so the observed signal in the nucleon
channel can be interpreted in terms of nucleon state with non-vanishing
coupling constant, possibly with an equal contribution from the
negative parity state. As was the case in the rho meson channel, the
repulsion observed in the delta channel can be explained in terms
of a chiral quark mass $m\simeq 85$ MeV.

\begin{table}
\begin{center}
\begin{tabular}{||ll|r|r|r|r|r|r||} \hline
  &  & RILM, $T=0$ & $f=0.25$ & $f=0.50$  & $f=0.75$ &
                     $f=0.95$ & $f=1.00$ \\ \hline
$\lambda^2_\pi$ & $[{\rm fm}^{-4}]$ &
   44$\pm$10 & 99$\pm$6  & 61$\pm$3 & 49$\pm$3 &
   22$\pm$4 & 1.8$\pm$0.3  \\
$\lambda^2_\delta$ & $[{\rm fm}^{-4}]$ &
   $< 0.1$ & $< 0.1$ &$< 0.1$ &$< 0.1$ &
   $< 0.1$ & 1.8$\pm$0.3  \\
$\lambda_\pi f_\pi$ & $[{\rm fm}^{-3}]$ &
   3.7$\pm$0.5 & 1.7$\pm$0.15 & 1.2$\pm$0.12 &
   0.8$\pm$0.06 &  0.3$\pm$0.06 &  $<$0.1\\
$c_\rho$ & $[{\rm fm}^{-3}]$   &
   6$\pm$1 & 1.7$\pm$0.24 &
   1.5$\pm$0.24 & 0.8$\pm$0.06 & 0.4$\pm$0.06 & $<$0.06\\
$ \beta_{N1}$ & $[{\rm fm}^{-7}]$   &
   80$\pm$17 & 150$\pm$26 &109$\pm$25  &
   63$\pm$25 &  25$\pm$15 & $<$10 \\
$\beta_{\Delta 1}$ & $[{\rm fm}^{-7}]$ &
   130$\pm$35 & 122$\pm$15 & 85$\pm$15 &
    63$\pm$13 & 29$\pm$10 & $<$8 \\
$\beta_{N2}$ & $[{\rm fm}^{-6}]$   &
    16$\pm$2  & & & & & 1.1$\pm$0.7 \\
$\beta_{N6}$ & $[{\rm fm}^{-7}]$  &
              &930$\pm$60 & 730$\pm$50 & 500$\pm$50
 & 190$\pm$40 &$<$20 \\ \hline
\end{tabular}
\end{center}
\caption{Coupling constants extracted from the parametrization of the
temporal correlation functions decsribed in the text, versus the fraction
of instanton molecules, $f$. The coupling constants are also defined in
the text, for more details see [9]. For comparison, we also show
our zero temperature results for the completely random ensemble ($f=0$). }
\end{table}

    Finally, the off-diagonal nucleon correlator $\Pi_6^N$ is shown in
in fig.2b. This correlator shows a very large signal at $f=0$ which
completely disappears as $f\rightarrow 1$. This is a clear indication
of $U_A(1)$ restoration and, most likely, parity doubling in the
nucleon spectrum. In this case the resonance contribution is $\beta_{N6}
=\lambda_{1N(\frac12^+)}\lambda_{2N(\frac12^+)}+\lambda_{1N(\frac12^-)}
\lambda_{1N(\frac12^-)}$, so $U_A(1)$ restoration implies a relation
between the various couplings.

    We have summarized our results for the coupling constants in table 1.
For comparison we have also included our results at zero temperature.
Although one should keep in mind that the resonance couplings at $T=0$
and $T=T_c$ are determined at different distances, a rough comparison
can still be made. Most channels simply show a gradual decrease of the
coupling constants, except for the pion and the nucleon, where the
couplings at $T=150$ MeV and $f=0.25$ are actually bigger than the
$T=0$ values. Most of this effect is probably an artefact due to the
use of a completely random ensemble at finite temperature. This ensemble
also gives a quark condensate which is somewhat larger than the $T=0$
value. We will study this point in more detail in a forthcoming
publication, where results from fully interacting ensembles will be
presented \cite{SSV_95b}. Comparing the squared pion coupling at $T=0$ to
its value at $f=1$ one finds that it drops by about a factor 20: assuming the
coupling to be proportional to $|\psi(0)|^2$, the probability to find
a quark and an antiquark at the same point, which is inversely proportional
to the volume, one concludes that the radius of the pion increases by
nearly a factor 3.

     In summary, we have presented an analysis of temporal correlation
functions for mesons and baryons in the instanton vacuum. In this model,
chiral symmetry restoration is due to the formation instanton-antiinstanton
molecules. In agreement with lattice results
\cite{BGK*94}, we have found significant non-perturbative effects for
scalar-pseudoscalar mesons even in the chirally symmetric phase.
Unfortunately, much better accuracy is needed to accurately determine
the mass or even the width of these states and settle the question
whether they are just some broad enhancement in the spectrum or fairly
narrow resonaces. We also find an attractive interaction in the
nucleon channel, while it is abesent in the case of the delta. It may
indicate that some hadronic states are not completely dissolved
at the phase transition, but survive as weakly bound $U(N_f)\times
U(N_f)$ chiral multiplets.

\newpage\noindent
{\Large\bf figure captions}\\ \\ \\
\underline{figure 1}
Isovector meson correlation functions in the temporal direction,
at $T=150$ MeV, normalized to free quark propagation. Fig.1(a) shows
the pseudoscalar (pion) correlator, (b) the scalar isovector (delta),
(c) the pseudoscalar-axialvector correlator and (d) the vector-tensor
one. All correlators are shown vor various fractions of instanton
molecules: open squares are $f=0.25$, open hexagons $f=0.50$,
skeletal stars $f=0.75$, crosses $f=0.95$ and open stars $f=1.00$.
\\ \\
\underline{figure 2}
Temporal correlation functions as in fig.1. Fig.2(a) shows the chiral
odd nucleon correlator $\Pi_1^N$, (b) the chiral even but $U_A(1)$
breaking correlator $\Pi_6^N$, (c) the chiral even nucleon and
delta correlators $\Pi_2^{N,\Delta}$ compared to the pion and rho
at $f=1$ only, and (d) the chiral odd delta correlator $\Pi_1^\Delta$.
\\ \\
\pagestyle{empty}
\newpage
\begin{figure}
\begin{center}
\leavevmode
\epsffile{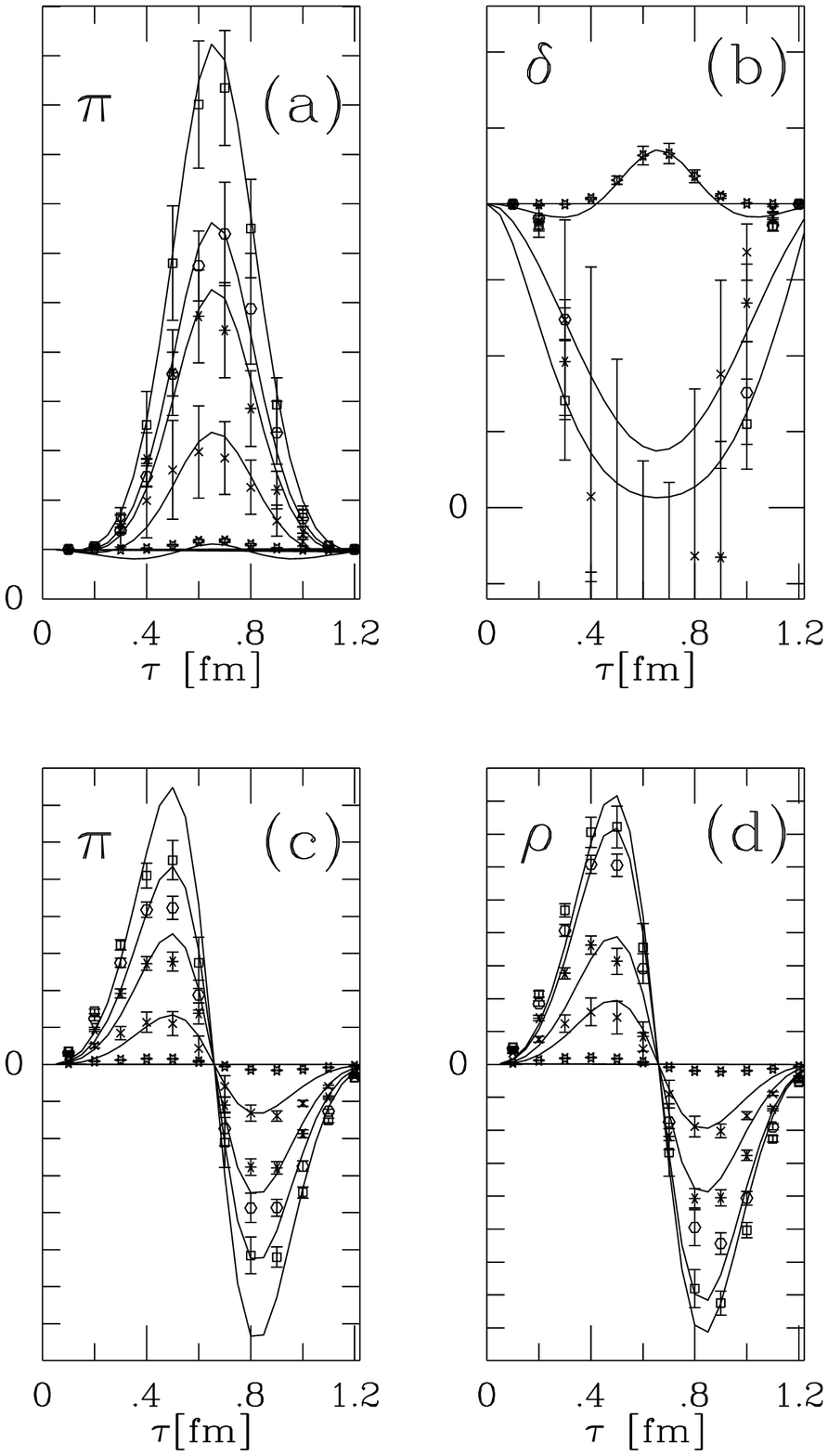}
\end{center}
\caption{}
\end{figure}
\vfill

\newpage
\begin{figure}
\begin{center}
\leavevmode
\epsffile{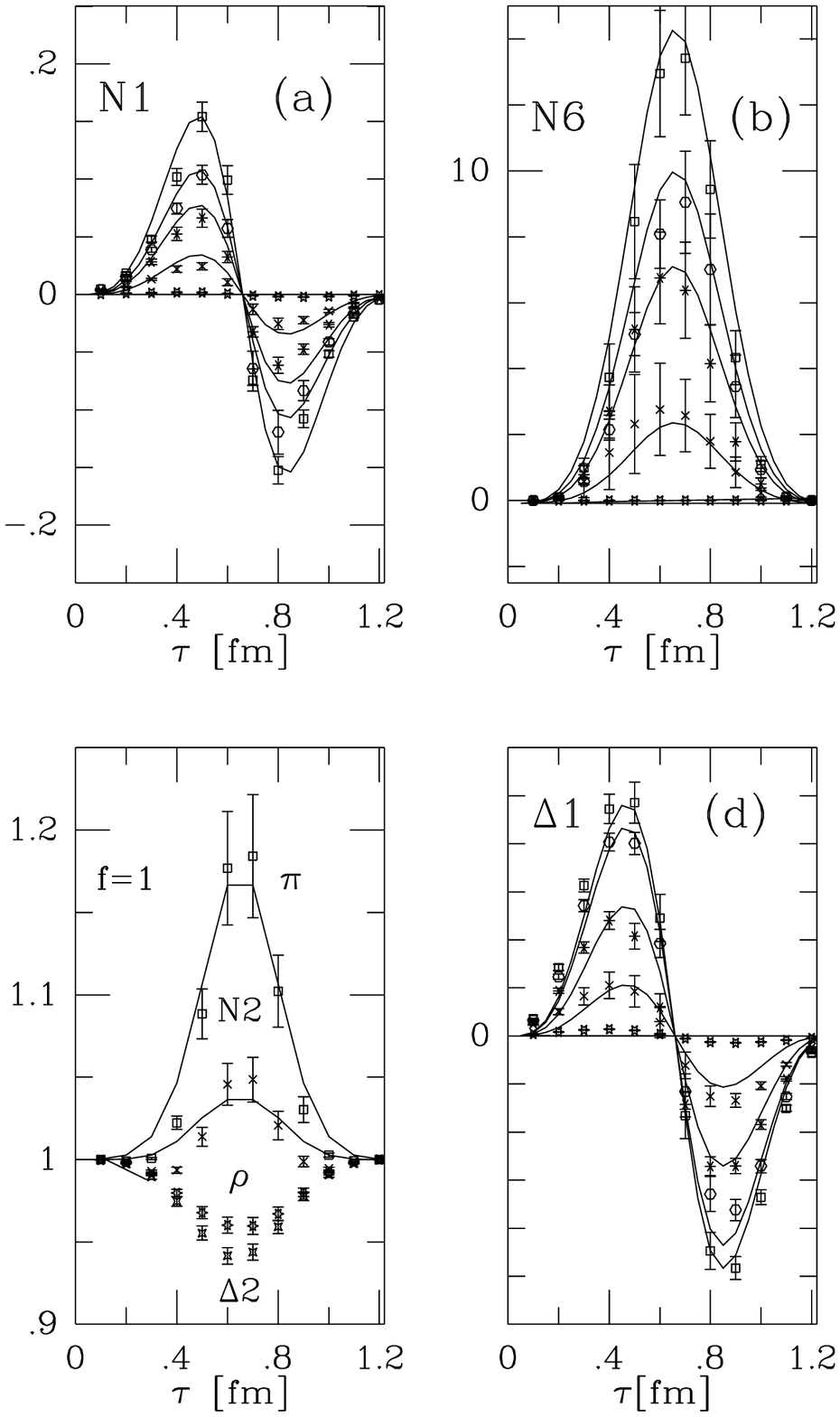}
\end{center}
\caption{}
\end{figure}
\vfill

\end{document}